%% file: draft_ZnSe_PRB_arxiv-v2.tex
\documentclass[aps,prb,showpacs,superscriptaddress,twocolumn]{revtex4-2}

\usepackage{natbib}
\usepackage{amsmath}
\usepackage{amssymb}
\usepackage{bm}
\usepackage{graphicx}
\usepackage{times}
\usepackage{wrapfig}
\usepackage{array}
\usepackage{color}
\usepackage{arydshln}
\usepackage{nicefrac}
\usepackage{upgreek}
\usepackage[normalem]{ulem}

\graphicspath{{figures/}}

\begin{document}
\title{Nondegenerate two-photon absorption in ZnSe: Experiment and theory}

\author{L.~Krauss-Kodytek}
\affiliation{Experimentelle Physik 2, Technische Universit\"at Dortmund, 44227 Dortmund, Germany}
\author{W.-R.~Hannes}
\author{T.~Meier}
\affiliation{Department of Physics, Paderborn University, Warburger Str. 100, 33098 Paderborn, Germany}
\author{C.~Ruppert}
\author{M.~Betz}
\affiliation{Experimentelle Physik 2, Technische Universit\"at Dortmund, 44227 Dortmund, Germany}

\date{\today}

\begin{abstract}
We experimentally and theoretically investigate the nondegenerate two-photon absorption coefficient $\beta(\omega_1,\omega_2)$ in the prototypical semiconductor ZnSe. In particular, we provide a comprehensive data set on the dependence of $\beta(\omega_1,\omega_2)$ on the nondegeneracy parameter $\omega_1/\omega_2$ with the total frequency sum $\omega_1+\omega_2$ kept constant. We find a substantial increase of the two-photon absorption strength with increasing $\omega_1/\omega_2$. In addition, different crystallographic orientations and polarization configurations are investigated. The nonlinear optical response is analyzed theoretically by evaluating the multiband semiconductor Bloch equations including inter- and intraband excitations in the length gauge. The band structure and the matrix elements are taken as an eight-band ${\bm k\cdot}{\bm p}$ model. The simulation results are in very good agreement with the experiment.
\end{abstract}


\maketitle	

\section{Introduction}

Semiconductors are characterized by large third-order optical nonlinearities, making them excellent materials for nonlinear photonic applications. These nonlinearities exist in both refractive and absorptive fashions and have been experi\-men\-tally studied and theoretically modelled in many materials. Among those phenomena, two-photon absorption (2PA) and its wavelength dependence are particularly well investigated~\cite{Stryland1985a,Sheik-Bahae1990}. Many previous studies are related to degenerate configurations where, e.g., the strength of the 2PA is studied in a $z$-scan experiment with one driving laser field. Less attention has been paid to the case of nondegenerate configurations even though they can occur in a variety of configurations. Nondegenerate 2PA is triggered by two laser fields at different frequencies. Some experiments have studied driving fields close to degeneracy~\cite{Bolger1993}, others $\omega$/2$\omega$ configurations~\cite{Sheik-Bahae1992}. Also situations with widely nondegenerate fields have attracted attention~\cite{Fishman2011,Cirloganu2011}. Here a massive enhancement of the two-photon absorption strength with increasing nondegeneracy parameter $\omega_1/\omega_2$ is observed.

Theoretical studies~\cite{Pidgeon1979,Vaidyanathan1980,Weiler1981,Wherrett1984} have suggested characteristic scaling laws for the spectral and bandgap dependence of the two-photon absorption coefficient $\beta$. 
While most of these dependencies are well established, the nondegenerate enhancement, which is best characterized by the dependence  on the ratio $\omega_1/\omega_2$ for a fixed sum frequency $\omega_1+\omega_2$, has been theoretically investigated mainly by rather simple two-band models so far~\cite{Aversa1994,Sheik-Bahae1994,Hannes2019}. 
A validation by a detailed comparison to experiment, meaning data sets for $\beta(\omega_1,\omega_2)$ with $\omega_1+\omega_2 = \text{const.}$, would be highly desirable. 

With this paper we fill this gap and provide several results for the nondegenerate two-photon absorption coefficient $\beta(\omega_1,\omega_2)$ for the prototypical semiconductor zinc selenide (ZnSe). 
We investigate configurations where the polarization of the two driving fields is either aligned parallel or perpendicular and additionally distinguish between different crystallographic orientations. We compare these results to numerical simulations of the two-photon absorption strength for this material, using the 8-band Kane model in combination with the semiconductor Bloch equations (SBE). Very good agreement with the experiment is achieved, without the use of any fitting parameter, for the magnitude of $\beta(\omega_1,\omega_2)$ and its dependence on $\omega_1/\omega_2$, the crystallographic direction, and the polarization direction.

\section{Experimental details}
\label{sec:experiment}

The experimental setup is based on a Ti:sapphire regenerative amplifier laser system (Coherent RegA 9000). It provides optical pulses with a central photon energy of 1.55\,eV, a pulse energy of $8\,\upmu$J and a pulse duration of $\approx60\,$fs at a repetition rate of 250\,kHz. Using a beam splitter a major fraction of these pulses is fed into an optical parametric amplifier (Coherent OPA 9400/9450) pumped by the second harmonic of these pulses. It emits signal and idler pulses with a fixed sum energy of $\hbar$$\omega_\mathrm{sig} + \hbar$$\omega_\mathrm{idl}= 3.1\,$eV. The corresponding individual tuning ranges are $\lambda_\mathrm{sig} = (510 - 800)\,$nm and $\lambda_\mathrm{idl} = (1855 - 800)\,$nm. Note that the configuration of $\lambda_\mathrm{sig} = \lambda_\mathrm{idl} = 800\,$nm is directly provided by the RegA system, but is included in the above tuning ranges. Thus, the OPA allows to systematically investigate frequency configurations $\omega_\mathrm{sig}/\omega_\mathrm{idl}$ from unity to 3.6. The pulse lengths and energies depend on the wavelength configuration and range from $\tau_\mathrm{sig}$ = $\tau_\mathrm{idl} = (70 - 460)\,$fs (full width at half maximum), $\varepsilon_\mathrm{sig} \le$ 140 nJ and $\varepsilon_\mathrm{idl} \le$ 40 nJ. Both beams are linearly polarized.  

The experimental setup is schematically shown in Fig.~1. In essence, we measure the strength of the 2PA in a pump-probe fashion with temporally overlapping signal and idler pulses. To this end, the idler pulse passes a motorized linear translation stage to adjust the time delay $\tau_\mathrm{d}$ between the two pulses. Afterwards, the signal and idler beam are superimposed on the sample. An angle of 20$^ \circ$ between the two beams allows us to easily separate them afterwards. Note that the angle within the sample (9$^ \circ$) is even smaller due to refraction and, thus, signal and idler directions can be assumed to be parallel inside the sample. A broadband $\lambda$/2-waveplate allows the polarization of the idler pulse to be rotated by 90$^ \circ$. In this way, we can compare 2PA measurements where the signal and idler pulses are polarized either parallel or perpendicular to each other. In the following, these two configurations will be referred to as co-polarized and cross-polarized. The idler pulse is blocked after the sample. The detection of the signal transmittance is performed with the lock-in technique referenced to a modulation of the idler beam with a mechanical chopper.

\begin{figure}
	\centering
	\includegraphics[width=1.\linewidth]{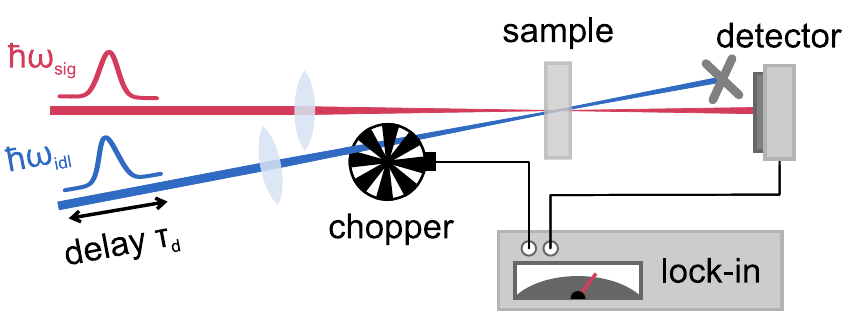}
	\caption{Experimental setup for the measurement of nondegenerate 2PA.}
	\label{fig:tpasetup}
\end{figure}

For the quantitative analysis of the two-photon absorption coefficient $\beta$, a precise knowledge of the spot diameters in the sample plane is important. To this end, we use the knife-edge-method with a titanium plate mounted in the sample plane acting as a knife. The peak irradiances required for the calculation of $\beta$ also depend on the pulse durations. To measure these pulse durations of the idler, we use a commercial autocorrelator (APE Mini TPA) that can be equipped with different detector units depending on the actual wavelength of the pulse. The signal pulses could also be temporally characterized here. However, the duration of the signal is easily determined by the data analysis described below.

We now turn towards the characterization of the samples. Zinc selenide (ZnSe) belongs to the group of II-VI-semiconductors with zincblende structure. The material is a direct semiconductor with a bandgap of $\varepsilon_\mathrm{g}$ = 2.7 eV at room temperature~\cite{Madelung1996}. As for the 2PA the sum energy of the involved photons needs to exceed the bandgap, the laser system in combination with the OPA, delivering pulses with $\hbar$$\omega_\mathrm{sig} + \hbar$$\omega_\mathrm{idl}= 3.1\,$eV, is perfectly suited for the investigation of ZnSe. The excess energy above the bandgap is $\Delta \varepsilon = 400\,$meV. We investigate ZnSe specimens with in both (100) and (110) orientation. The sample thicknesses are $42\,\upmu$m for the (100) oriented and $37\,\upmu$m for the (110) oriented sample, respectively. These thicknesses are achieved by mechanical grinding of commercial wafers. The thicknesses are chosen to ensure homogeneous excitation conditions across the sample depth. The orientation of the light polarizations relative to the crystallographic axes is shown in Fig.~2 for both samples.

\begin{figure}
	\centering
	\includegraphics[width=.8\linewidth]{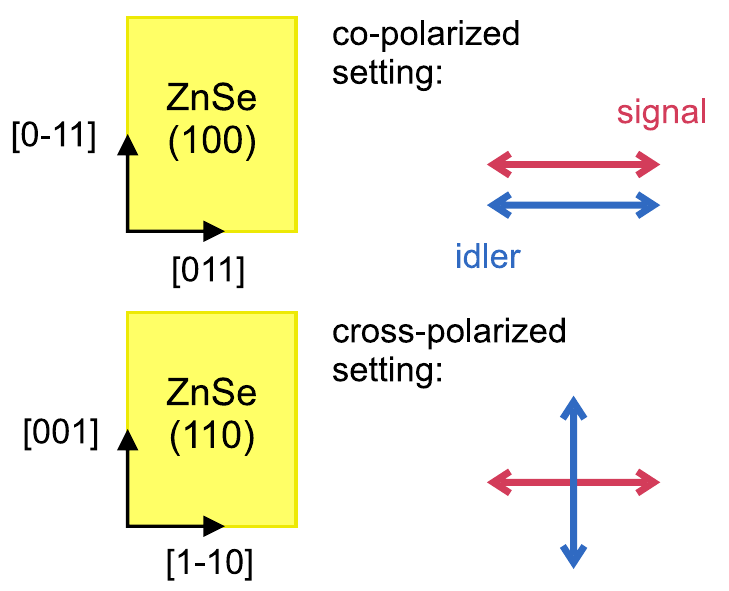}
	\caption{Orientation of the signal and idler polarizations with respect to the crystallographic axes for the (100) and (110) ZnSe sample.}
	\label{fig:settings}
\end{figure}

\section{Theoretical model}
\label{sec:theory}

One of the simplest band-structure models suitable for investigating 2PA in zincblende semiconductors for different relative polarization directions between the two pulses is the 8-band Kane model with nonparabolicity~\cite{Kane1957}. 
This model includes the lowest conduction band and the light-hole, heavy-hole and split-off valence bands. It was used in Refs.~\cite{Hutchings1992,Bolger1993} and provided an improved frequency-dependence of the (nondegenerate) $\beta$ over the two-band models used in earlier studies~\cite{Pidgeon1979,Vaidyanathan1980,Weiler1981,Wherrett1984}.
However, the model is still isotropic and thus does not capture the dependence on the crystal orientation. 
From measurements, such an anisotropy has been known to exist for a long time. 
In Refs.~\cite{Hutchings1994,Dvorak1994} it was shown that the anisotropy of the (degenerate) $\beta$ is mainly due to the influence of higher conduction bands, not by serving as intermediate states (at least for the considered frequencies), but by inducing warping of the valence bands.
For the numerical approach described here we also use an 8-band Kane model, but include the interaction with higher conduction bands in second-order perturbation theory \cite{Winkler2003} which leads to higher-order terms in the band energies $\varepsilon_k$. 
Where available, we compare our findings with the predictions made by two-band models in 
transition rate~\cite{Said1992,Sheik-Bahae1991,Sheik-Bahae1994} or susceptibility~\cite{Aversa1994,Sipe1993,Aversa1995} approaches.

A major difficulty in the combination of dynamical equations in the length gauge with rather realistic band-structure models is the random gauge of the numerically obtained eigenfunctions/matrix elements at each ${\bm k}$-point. We build on earlier works~\cite{Souza2004,Vanderbilt2018,Virk2007,Thong2021}, to reconstruct a globally smooth gauge in the modeled region of ${\bm k}$-space. This is achieved by applying a parallel transport gauge (PTG), where additionally the lines in ${\bm k}$-space are connected via a common starting point. With this procedure further gauge transformations during the numerical time evolution are not required for any polarization setting. Since we transform the gauge such that connected bands are mixed~\cite{Virk2007} (not only where they are exactly or nearly degenerate), the same transformation needs to be applied to all other matrices (denoted by the hat symbol; note that the matrices other than $\bm{\xi_k}$ transform as usual operators). The dynamical equation for the single particle density matrix $\rho_{\bm k}$ can be expressed (independently of the gauge) as
\begin{align}\label{eq:SBE_length-gauge}
\frac{\partial}{\partial t} \hat{\rho}_{\bm k}
&= 
\frac{i}{\hbar} \big[ \hat{h}_{\bm k} - e\bm{E}(t) \cdot 
\hat{\bm \xi}_{\bm k},\hat{\rho}_{\bm{k}}\big]
+
\frac{1}{\hbar}  {e\bm{E}}(t) \cdot 
{\nabla}_{\bm k} \hat{\rho}_{\bm k} .
\end{align}
The optical response is determined by the local spatial average of the mean microscopic current density,
\begin{align}
\bm{J} &= -e \frac{1}{L^3} \sum_{\bm{k}} \mathrm{Tr} ( \hat{\rho}_{\bm k}\hat{\bm{v}}_{\bm k} ).
\end{align}
The input matrices ($h_{\bm k},\bm{\xi_k},\bm{v_k}$) 
to Eqs.~(1) and (2) are ground state properties defined as follows. 
In the original gauge, $h_{\bm k}$ is the diagonal matrix of the single-particle band energies $\varepsilon_k$. 
The velocity matrix $\bm{v_k}$ is defined as
\begin{align}
\hbar\hat{\bm{v}}_{\bm k} = {\nabla}_{\bm k} \hat{h}_{\bm k} + i \big[\hat{h}_{\bm k},\hat{\bm \xi}_{\bm{k}}\big],
\end{align}
where in the original gauge the second term corresponds to the interband contribution. Note that Eqs.~(2) and (3) hold for both the original gauge and our PTG, which leaves $h_{\bm k}$ smooth. 
The transition dipoles 
$e\bm{\xi}_{\bm{k}}^{\lambda\lambda'}$
are obtained as matrix elements of the position operator from the periodic parts of the Bloch states,
\begin{align}
\bm{\xi}_{\bm{k}}^{\lambda\lambda'} &\equiv i \langle u^\lambda_{\bm{k}} | \nabla_{\bm{k}} | u^{\lambda'}_{\bm{k}} \rangle.
\end{align}
This also holds for the intraband ($\lambda=\lambda'$) elements which, however, requires a smooth gauge. The gradient terms in Eqs.~(1), (3), and (4) are evaluated numerically as symmetrized difference quotient in the discretized ${\bm k}$-space.

Our approach is fully consistent in the sense that different gauges lead to the same result for the optical response. Eq.~(1) corresponds to the SBE in the length gauge. In this study, the commonly used phenomenological dephasing and relaxation terms are not included, since the corresponding time scales are expected to be longer than the pulse length in the experiment and because the effect of dephasing is essentially limited to a broadening of spectral features. Likewise, many-body effects are not expected to dominate as the sum frequencies are sufficiently high above the band gap and are thus omitted as in previous studies on 2PA. We are interested in 2PA resulting from the simultaneous absorption of one signal and one idler photon. This process is isolated from other responses via a directional order expansion in the two electric fields (where $\bm{E}$(t) in Eq.~(1) is the sum of both, i.e., $\bm{E}$(t)$=$$\bm{E}_\mathrm{sig}$(t) $+$$\bm{E}_\mathrm{idl}$(t))~\cite{Hannes2019}. The absorbed two-photon energy is given by the time-integrated product of the signal-directed third-order current $\tilde{{\bm J}}\equiv\bm{J}^{(3)(1|0)}$ with the signal electric field. Normalization with respect to the incident intensities leads to the 2PA coefficient~\cite{Hannes2019,Mahr1975}
\begin{align}
\label{eq:PrAbs_pulse}
\beta(\omega_1,\omega_2) &=
\frac{1}{\epsilon_0^2 c^2}
\frac{1}{n_\mathrm{sig} n_\mathrm{idl}}
\frac{\displaystyle\int \mathrm{d}t\,
	\langle\tilde{{\bm J}}(t) \cdot \tilde{{\bm E}}_\mathrm{sig}^*(t)\rangle}{\displaystyle\int \mathrm{d}t \hat{E}_\mathrm{sig}^2(t) \hat{E}_\mathrm{idl}^{2}(t)} ,
\end{align}
where $n_\mathrm{pulse} \equiv n(\omega_\mathrm{pulse})$ is the linear refractive index at the frequency of the (signal or idler) pulse.
Since the variation of $n_\mathrm{sig} n_\mathrm{idl}$ in the considered parameter range (with fixed sum frequency) is small, we approximate the frequency-dependent refractive index $n(\omega)$ by~\cite{Marple1964}
\begin{align}
	n^2 = A + \frac{B\lambda^2}{\lambda^2-C^2},
\end{align}
where $\lambda$ is the wavelength in $\upmu\text{m}$, and the parameters for ZnSe have been experimentally determined as: $A=4.00$, $B=1.90$, and $C^2=0.113$.~\cite{Marple1964}
Thus, in our case the factor $n_\mathrm{sig}n_\mathrm{idl}$ increases by about 6\% when changing the nondegeneracy parameter $\omega_1/\omega_2$ from unity to 4 (corresponding to the range considered in Sec.~\ref{sec:results}), with the sum frequency fixed at 3.1~eV.

In the numerical simulations, 
signal and idler pulse are assumed to be co-propagating along the sample orientations given in Fig.~2.
We use pulses with a Gaussian envelope of duration $\tau_\mathrm{sig} = \tau_\mathrm{idl} 
\approx 29.44\,$fs
(full width at half maximum of the pulse intensity). 
Since the computational expense increases higher than linearly with the pulse durations (due to demanding finer ${\bm k}$- and $t$-resolutions and a longer duration of the integration), we have refrained from simulating longer pulses to keep the numerical effort within resaonable limits. However, in our case of fixed sum frequency, the higher spectral resolution provided by longer pulses is not required since it does not significantly alter the shown spectra (as confirmed by simulations using dimensionally reduced $k$-grids). 
As mentioned in Sec.~\ref{sec:results}, the evaluation of experiments with different pulse durations also provided practically identical 2PA coefficients.

The band-structure input is provided by the 8-band Kane model with remote band contributions, using the parameters tabulated in Ref.~\cite{[][{. The explicit matrix form of the 8-band Kane model is given in Table C.8, where we neglect all terms proportional to Kane's parameters ($B_{8v}^\pm$ and $B_{7v}$). Table D.1 lists bulk band structure parameters. The effective mass and Luttinger parameters have to be reduced according to Table C.9.}]Winkler2003}. In order to investigate different crystallographic orientation, a rotation of the crystal is applied for the (110) sample, such that the quantization axis of the total angular momentum (the $z$-axis in the ${\bm k\cdot}{\bm p}$ theory) corresponds to the propagation direction. The simulated 3D ${\bm k}$-space is resolved by a Cartesian grid of sufficiently large side length (here 0.2  $\pi/a_0$, where  $a_0$ is the lattice constant), with the $\Gamma$-point in the center. For the chosen resonance frequency of $\hbar$$\omega_\mathrm{sig}$ + $\hbar$$\omega_\mathrm{idl}$ about 400~meV above the band gap, the accuracy of the 8-band ${\bm k\cdot}{\bm p}$ model may be limited. We can, however, estimate that this together with the limited resolution and the short pulse duration causes an uncertainty of the obtained values for $\beta$ of roughly 15\%.

The perturbation expansion of the SBE (1) can be solved analytically in the cw-limit, which has been demonstrated for the $\chi^{(3)}$-response within a simple two-band model with ${\bm k}$-independent dipole matrix element in Ref.~\cite{Hannes2019}. That approach results in a scaling $\beta \sim  n_\mathrm{sig}^{-1} n_\mathrm{idl}^{-1}  \omega_\mathrm{sig}^{-3}\omega_\mathrm{idl}^{-4}(\omega_\mathrm{sig}^2+\omega_\mathrm{idl}^2)^2$ (not shown), which corresponds to a slightly stronger increase of the 2PA with the nondegeneracy parameter than our present numerical results with ${\bm k}$-dependent dipole matrix elements and than the experimental results. If we alter the model of Ref.~\cite{Hannes2019}
by the following ${\bm k}$-dependence of the transition dipole, $\xi_{\bm k} = \xi_{0} (\omega_{0}/\omega_{\bm k})^2$, we find $\beta \sim n_\mathrm{sig}^{-1} n_\mathrm{idl}^{-1} \omega_\mathrm{sig}^{-3}\omega_\mathrm{idl}^{-4}$, which for fixed sum frequency is the same frequency dependence as obtained from the approaches of Refs.~\cite{Sheik-Bahae1991} and \cite{Aversa1994} and also agrees well with the results shown below (see curves labeled `2-band' in Fig.~4). 
Identifying a similar scaling rule for the cross-polarized case is less trivial since it would require analyzing the magnitude and directions of the complex transition dipoles in a more realistic band-structure model including spin.

\section{Results and discussion}
\label{sec:results}

To experimentally investigate the nondegenerate 2PA coefficient $\beta(\omega_\mathrm{sig},\omega_\mathrm{idl})$, depending on the frequency ratio $\omega_\mathrm{sig}/\omega_\mathrm{idl}$ of the two involved photons, we measure various data points within the range of possible frequency combinations of signal and idler photons. Our experiment and the detection scheme is designed to measure 2PA whereby one photon from each beam triggers 2PA. In order to suppress degenerate 2PA with two photons from the same beam, the energies of the signal and idler photons are chosen to be $\varepsilon_g/2 <\hbar\omega_\mathrm{sig}<\varepsilon_g$ and $\hbar\omega_\mathrm{idl}<\varepsilon_g/2$. With this choice the idler alone cannot trigger 2PA. The idler is chosen as the pump beam in our pump-probe approach. The signal acts as the probe beam. Its intensity is much weaker than the idler such that also the signal alone does not induce significant 2PA. The spot sizes of the pulses are adjusted that the idler spot is at least 2.5 times larger than the signal spot. 
As a result, the intensity of the idler across the signal spot is practically identical to its peak intensity such that spatial averaging is not necessary for the extraction of $\beta$.

\begin{figure}
	\centering
	\includegraphics[]{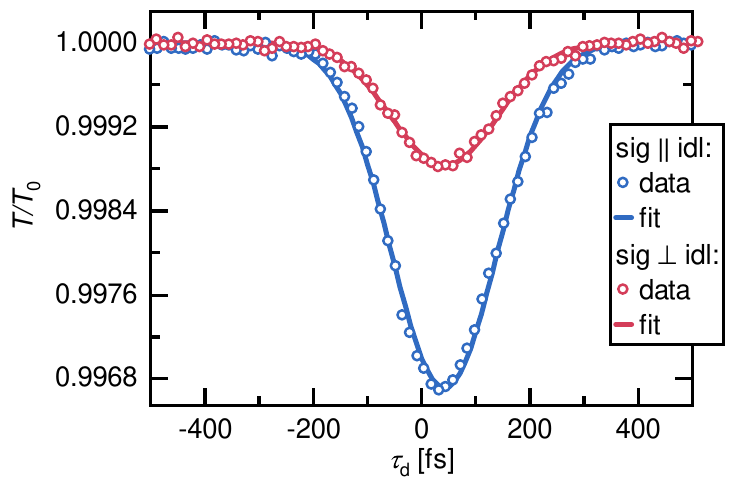}
	\caption{Exemplary pump-probe transients for a co- and cross-polarized 2PA measurement at $\lambda_\mathrm{sig} = 570\,$nm and $\lambda_\mathrm{idl} = 1341\,$nm for the (100) oriented ZnSe sample. The blue (red) data correspond to the co- (cross-) polarized configuration. Solid lines represent a fit according to the model by Negres \textit{et~al.}~\cite{Negres2002}.}
	\label{fig:transients}
\end{figure}

Fig.~3 shows an exemplary pump-probe transient for a co- and cross-polarized measurement at a wavelength combination of $\lambda_\mathrm{sig} = 570\,$nm and $\lambda_\mathrm{idl} = 1341\, $nm for the (100) oriented ZnSe sample. The signal and idler pulses have Gaussian temporal pulse shapes and the measured idler pulse length is $\tau_\mathrm{idl} = 98.2\,$fs (full width at half maximum). 
While our pulse durations are somewhat longer than the ones in the simulation, we do not expect an impact for the resulting 2PA coefficients.
In fact, by using different alignments of the OPA, we have repeated the experiment with different pulse durations for one wavelength combination.
After incorporating the actual pulse durations into the data analysis, we found practically identical 2PA coefficients

To evaluate the 2PA coefficient $\beta$ we use the theoretical approach for the normalized signal transmittance of Negres \textit{et~al.}~\cite{Negres2002}. In short, the results of Negres \textit{et~al.} establish an analytical formula for the shape of the pump-probe transients for our experimental configuration. It assumes pulses with a Gaussian temporal envelope. It takes into account the well- known group velocity mismatch of signal and idler pulses. While the approach of Negres \textit{et~al.} in principle takes linear absorption into account as well, we set this linear absorption to zero. 
This assumption is valid because the linear absorption across the bandgap is supressed by the chosen photon energies ($<\epsilon_\text{g}$). Free carrier absorption can be excluded as the pump-probe traces do not show transient absorption beyond the temporal overlap of pump and probe pulse.
A fit of this model to the experimental transients reveals the co- and cross-polarized 2PA absorption coefficients $\beta_\parallel$, $\beta_\perp$, and, in addition, the duration of the signal pulse. Two examples for such a fit are displayed in Fig.~3 and show excellent agreement between the model and the measurement.

For the specific wavelength combination presented in Fig.~3, the fit represents curves with $\beta_\parallel = 12.93\,_{-1.4}^{+1.6}\,$cm/GW with $\tau_\mathrm{sig}= (120.7 \pm 0.9)\,$fs and $\beta_\perp = 4.77\,_{-0.5}^{+0.6}\,$cm/GW with $\tau_\mathrm{sig}= (120.1 \pm 1.7)\,$fs. The pulse lengths are given as the full width at half maximum. The spot sizes for this example are $A_\text{sig}=2100\,\upmu\text{m}^3$ and $A_\text{idl}=30100\,\upmu\text{m}^3$. We estimate the errors for $\beta$ by varying the input parameters within the measurement inaccuracies. Since all the pump-probe data sets look rather comparable and the analysis follows analogously to the above procedure, we now restrict the presentation to the values for the two-photon absorption coefficient $\beta$ and its dependence on the nondegeneracy, polarization, and crystal orientation.

\begin{figure}
	\centering
	\includegraphics[width=1.\linewidth]{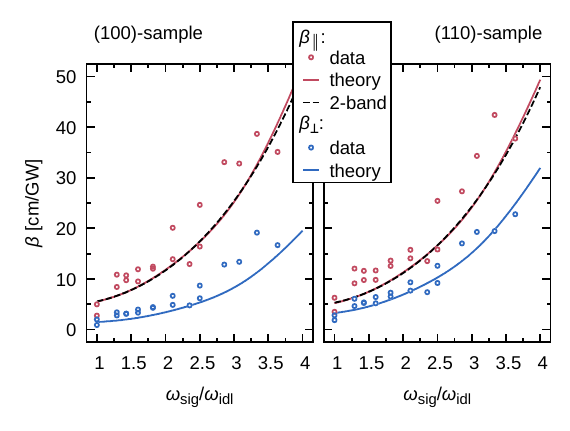}
	\caption{2PA coefficients of ZnSe as a function of the nondegeneracy parameter. The corresponding polarization settings for co- and cross-polarized cases and each crystallographic orientation are shown in Fig.~2. The curves labeled `theory' are obtained from simulations with $\tau=50\,$fs pulses. The curves labeled `2-band' correspond to $\beta_ \parallel \sim n_\mathrm{sig}^{-1} n_\mathrm{idl}^{-1} \omega_\mathrm{sig}^{-3}\omega_\mathrm{idl}^{-4}$, where the prefactor has been chosen for best fit with the numerical results.}
	\label{fig:beta}
\end{figure}

We now turn to the comparison of the experimentally observed 2PA coefficients for various nondegeneracy parameters $\omega_\mathrm{sig}/\omega_\mathrm{idl}$ and the two crystallographic orientations. As seen in Fig.~4, the values for $\beta_ \parallel(\omega_\mathrm{sig}/\omega_\mathrm{idl})$ and $\beta_\perp(\omega_\mathrm{sig}/\omega_\mathrm{idl})$ increase with the nondegeneracy parameter. Specifically, we find an approximately fivefold increase of $\beta$ when comparing configurations with $\omega_\mathrm{sig}/\omega_\mathrm{idl} = 3.6$ to the degenerate case $\omega_\mathrm{sig}/\omega_\mathrm{idl} = 1$. Note that the scatter of the experimental data also increases with $\omega_\mathrm{sig}/\omega_\mathrm{idl}$ because the noise of the optical parametric amplifier increases for widely nondegenerate operation. The increase of the two-photon absorption strength with higher degrees of nondegeneracy has been theoretically predicted in several studies~\cite{Hannes2019,Aversa1994,Sheik-Bahae1994}. When comparing the two crystallographic orientations, we find very similar magnitudes for $\beta_ \parallel(\omega_\mathrm{sig}/\omega_\mathrm{idl})$ at the same frequency combinations. In contrast, the values for $\beta_\perp(\omega_\mathrm{sig}/\omega_\mathrm{idl})$ for the (110)-oriented sample are consistently higher than for the (100)-oriented sample. We will elaborate further on this fact below.

It is also instructive to compare our results to previous studies of 2PA in bulk ZnSe. So far, comparable results are only available for de\-ge\-ne\-rate, co-polarized configurations and 800\,nm pulses. In this case, we find average $\beta$-values of 3.9\,cm/GW for the (100)-sample and 4.9\,cm/GW for the (110)-sample. Our experimental results agree well with the value of 3.5\,cm/GW independently reported in literature by Balu \textit{et~al.}~\cite{Balu2008} and Fishman \textit{et~al.}~\cite{Fishman2011}. Balu et al. used a polycrystalline ZnSe specimen. Fishman et al. do not give particular specifications of their sample.

Fig.~4 also contains the theoretical results for the same crystallographic orientations and polarization configurations. Given the uncertainties in both experiment and theory, the agreement for all four settings is surprisingly good. Without using any fitting parameter, the theory yields a very similar behavior of $\beta$ as a function of the nondegeneracy parameter, and even its absolute magnitudes are very close to the measured ones.

\begin{figure}
	\centering
	\includegraphics[width=1.\linewidth]{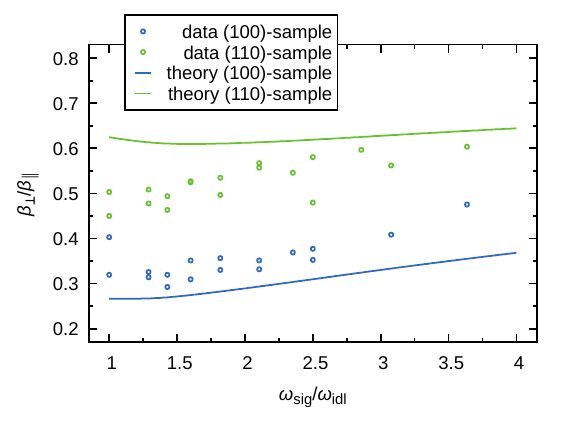}
	\caption{Ratio between cross- and co-polarized 2PA. Same data (experiment and theory) as in Fig.~4.}
	\label{fig:betaratio}
\end{figure}

We finally elaborate on the anisotropy of $\beta$ in terms of the different polarization configurations and its dependence on the nondegeneracy. A useful parameter to characterize the polarization anisotropy is the ratio $\beta_\perp/\beta_ \parallel$ for co- and cross-polarized signal and idler beams. These results are summarized in Fig.~5. The first observation is that the values of $\beta_\perp/\beta_ \parallel$ for the (100) oriented sample are clearly smaller than for the (110) sample. More surprisingly, we find a significant increase of $\beta_\perp/\beta_ \parallel$ with increasing $\omega_\mathrm{sig}/\omega_\mathrm{idl}$. This result has not been reported in any previous study on nondegenerate 2PA. An advantage of inspecting this ratio is its lower noise due to cancelling out of certain error sources in the original quantities. 

The numerical results from the 8-band Kane model lie within the experimental uncertainties. The experimental result for $\beta_\perp/\beta_ \parallel$ might be slightly overestimated due to the oblique idler propagation in the nominally co-polarized case. As discussed above, the numerical results come with their own uncertainty - in particular, the slight decrease of $\beta_\perp/\beta_ \parallel$ in the vicinity of the degenerate case might result from the finite resolution in ${\bm k}$-space. Otherwise, the properties of $\beta_\perp/\beta_ \parallel$ found in the experiment as described above are well captured by our simulations. 

To explain the microscopic origin of this behavior is complicated in the considered model, since
the 8$\times$8-Kane Hamiltonian with remote-band contributions cannot be diagonalized analytically.
However, we have evaluated a simpler model, in particular an anisotropic two-band model with complex ${\bm k}$-independent transition dipoles, obtained from zone-center wave functions~\cite{Hannes2020}. In this model, $\beta_\perp$ is given by $\beta_ \parallel$ with two out of four  excitation paths dropped for symmetry reasons. This results in a different frequency dependence for both cases and in the following dependence of the ratio on the nondegeneracy parameter,
\begin{align}
\frac{\beta_\perp}{\beta_\parallel} &= \frac{1+(\omega_1/\omega_2)^4}{[1+(\omega_1/\omega_2)^2]^2},
\end{align}
which equals $1/2$ for the degenerate case and increases towards unity for the strongly nondegenerate case.
In comparison with the realistic model, this increase is too strong as a consequence of the neglected ${\bm k}$-dependence of transiton dipoles. 

\section{Conclusions}
\label{sec:conclusions}

In summary, we have carefully investigated the nondegenerate two-photon absorption coefficient $\beta(\omega_1,\omega_2)$ in the prototypical semiconductor ZnSe for a fixed sum frequency $\omega_1+\omega_2$. We find a substantial increase of the two-photon absorption strength with increasing $\omega_1/\omega_2$ as predicted by different theoretical approaches. Specifically, we find an about fivefold increase of $\beta(\omega_1,\omega_2)$ for $\omega_1/\omega_2= 3.5$  when compared to the degenerate case. A numerical model based on eight-band ${\bm k\cdot}{\bm p}$ calculations and semiconductor Bloch equations including inter- and intraband excitations agrees quantitatively with the experiment in terms of the dispersion of $\beta(\omega_1,\omega_2)$ as well as the crystalline and polarization anisotropy. The results are important for a more detailed understanding of nonlinear two-color interactions in semiconductors and their applications in nonlinear photonics.

\acknowledgments
We acknowledge funding from the Deutsche Forschungsgemeinschaft (DFG) in the framework of the CRC – TRR142 (grant No. 231447078, project A07).
We also thank the PC$^2$ (Paderborn Center for Parallel Computing) for providing computing time.


\input{draft_ZnSe_PRB_RESUB_v1.2.bbl}

\end{document}

%% file: draft_ZnSe_PRB_RESUB_v1.2.bbl
%